

\input phyzzx
\overfullrule=0pt
\hsize=6.5truein
\vsize=9.0truein
\voffset=-0.1truein
\hoffset=-0.1truein

%
%
\def\alphat{\tilde \alpha}
\def\BH{\hbox{\ninerm BH}}
\def\eff{\hbox{\ninerm eff}}
\def\evap{\hbox{\ninerm evap}}

\def\MAX{\hbox{\ninerm MAX}}
\def\mirror{\hbox{\ninerm mirror}}
\def\outer{\hbox{\ninerm outer}}
\def\pa{\partial}
\def\Planck{\hbox{\ninerm Planck}}
\def\Rvec{\vec R}
\def\xvec{\vec x}
\def\Xvec{\vec X}

%
%
\rightline{SU-ITP-94-35}
\rightline{October 1994}
\rightline{hep-th/9410074}

\vfill

%
%
\title{Black Holes, Interactions, and Strings}

\vfill

%
%
\author{Leonard Susskind\foot{susskind@dormouse.stanford.edu}
and John Uglum\foot{john@dormouse.stanford.edu}}

\vfill

\address{Department of Physics \break Stanford University, Stanford,
CA
94305-4060}

\vfill

%
%
\abstract{\singlespace We give some examples in which neglecting the
interactions between particles or truncating the description of a
black hole to the spherically symmetric mode leads to unphysical
results.  The restoration of the interactions and higher angular
momentum modes resolves these problems.  It is argued that
mathematical consistency of the description of black holes in the
Schwarzschild coordinate system requires that we neither truncate the
theory nor ignore the interactions.  We present two hypotheses on how
matter must behave under large Lorentz boosts in order for black
holes to be consistent with quantum mechanics.  Finally, we argue
that string theory exhibits these properties.}

\vfill

{\it Talk presented at the PASCOS meeting in Syracuse, New York, May
1994.}

\vfill

%
%
PACS categories: 04.70.Dy, 04.60.Ds, 11.25.Mj, 97.60.Lf
\vfill\endpage

%
%

\REF\hawk{S.~W.~Hawking \journal Comm. Math. Phys. & 43 (75) 199.}

\REF\jac{Attempts to remove any reference to high frequencies in the
calculation have been made; see {\it e. g. \/} T.~Jacobson \journal
Phys. Rev. & D48 (93) 728.}

\REF\page{D.~N.~Page \journal Phys. Rev. & D13 (76) 198.}

\REF\CGHS{C.~G.~Callan, S.~B.~Giddings, J.~A.~Harvey, and
A.~Strominger \journal  Phys. Rev. & D45 (92) 1005.}

\REF\hawkgid{S.~W.~Hawking \journal Phys. Rev. & D14 (76) 2460; \hfil
\break
S.~B.~Giddings and W.~M.~Nelson \journal Phys. Rev. & D46 (92) 2486.}

\REF\stu{L.~Susskind, L.~Thorlacius, and J.~Uglum \journal Phys. Rev.
& D48 (93) 3743.}

\REF\thooft{G.~'t~Hooft \journal Nucl. Phys. & B256 (85) 727.}

\REF\witat{J.~J.~Atick and E.~Witten \journal Nucl. Phys. & B310 (88)
291.}

\REF\suss{L.~Susskind, {\it Some Speculations About Black Hole
Entropy in String Theory,} Rutgers University preprint RU-93-44,
August 1993, hep-th/9309145.}

\REF\su{L.~Susskind and J.~Uglum \journal Phys. Rev. & D50 (94)
2700.}

\REF\dab{A.~Dabholkar, {\it Quantum Corrections to Black Hole Entropy
in String Theory, \/} Caltech preprint CALT-68-1953, September 1994,
hep-th/9409158.}

\REF\kks{M.~Karliner, I.~Klebanov, and L.~Susskind \journal Int.
Jour. Mod. Phys. & A3 (88) 1981.}

\REF\mpt{A.~Mezhlumian, A.~Peet, and L.~Thorlacius \journal Phys.
Rev. & D50 (94) 2725.}

\REF\mempar{K.~S.~Thorne, R.~H.~Price, and D.~A.~MacDonald, {\it
Black
Holes:  The Membrane Paradigm},  Yale University Press, 1986, and
references therein.}

\REF\ggrt{P.~Goddard, J.~Goldstone, C.~Rebbi, and C.~Thorn \journal
Nucl. Phys. & B56 (73) 109.}

\REF\susstwo{L.~Susskind, {\it The World as a Hologram, \/} Stanford
University preprint SU-ITP-94-33, September 1994, hep-th/9409089.}

%
%

%
%
\chapter{Introduction}

The vast majority of work that has been done on the subject of black
hole evaporation in 3+1 dimensions, beginning with the seminal work
of Hawking [\hawk ], has relied on the approximation of free fields
propagating in the fixed background black hole geometry.  There is a
good reason for this, since solving an interacting quantum field
theory, even in Minkowski space, is extremely complicated.  Indeed,
there are certain calculations for which the free field approximation
gives a perfectly sensible answer.  As an example, recall that the
Euclidean continuation of the exterior Schwarzschild geometry for a
black hole of mass $M$ is periodic in the Euclidean time variable
$\Theta$ with period $8\pi MG$ ($G$ is the Newton constant).
Therefore, the Euclidean Green functions of any quantum field theory
on this background, interacting or not, will have this periodicity.
This shows that the only static state of the system is a thermal
state at the Hawking temperature $T_H = {1 \over {8 \pi MG}}$.  In
particular, since this holds for free field Green functions, we see
that free field theory is sufficient to get the basic thermodynamics
of the system correct.

However, the free field approximation leads to puzzling conclusions
for certain other questions.  For instance, the free field
approximation tells us that the black hole is in thermal equilibrium
at the Hawking temperature.  However, for ordinary systems one
expects to achieve thermal equilibrium only if there are interactions
present.  One wonders how a black hole could circumvent this.  A
second example of the inadequacy of the free field approximation is
that modes of arbitrarily high frequency, much higher than the Planck
mass, appear in the calculation of the properties of the Hawking
radiation.  Since we have no knowledge of physics beyond the Planck
scale, the calculation is suspect [\jac ].

A further approximation is often invoked, in which the system is
truncated to include only spherically symmetric modes.  Since almost
all of the escaping Hawking particles carry little or no angular
momentum [\page ], and since, in the absence of interactions, the
different angular momentum modes are decoupled, it is often argued
that these higher angular momentum modes are irrelevant to the
properties of the Hawking radiation.  Indeed, the spherically
symmetric description of a Schwarzschild black hole has been elevated
from the status of an approximation to that of an independent 1+1
dimensional mathematical model [\CGHS ].  As is well known, however,
the resulting description of the details of the Hawking radiation
leads to paradoxes and inconsistencies with quantum theory [\hawkgid
].

In the following we will give some examples of situations in which
neglecting interactions and/or truncating the theory to only the
spherically symmetric modes leads to unphysical results.  We will
then show how including the higher angular momentum modes and the
interactions resolves these problems.  We emphasize that the
mathematical consistency of the description of the black hole in the
Schwarzschild coordinate system requires that we not truncate the
theory.  Moreover, we will argue that for black holes to be
consistent with quantum theory matter must have very specific
properties under large Lorentz boosts.  Finally, we will see that
fundamental strings exhibit some of the necessary properties.

Let us begin by examining some consequences of truncating the theory
to the $s-$wave sector.  In the Schwarzschild coordinate chart
$(t,r,\theta,\varphi)$ the line element of the exterior Schwarzschild
geometry has the form
$$
ds^2 = - \left ( 1 - {{2MG} \over r} \right ) dt^2 + \left ( 1 -
{{2MG} \over r} \right )^{-1} dr^2 + r^2 \left ( d\theta^2 + \sin^2
(\theta ) d\varphi^2 \right ) \>,
\eqn\linelement
$$
where the horizon is at $r = 2MG$.  The entropy of the black hole is
given by the Bekenstein-Hawking formula
$$
S_{\BH} = {A \over {4G}} \>,
\eqn\BHent
$$
where $A = 16\pi M^2 G^2$ is the area of the horizon.  For ordinary
systems, the degrees of freedom that account for the entropy of a hot
system are also those which thermalize, store, and eventually reemit
any information which may have been absorbed by the system.  Later we
will discuss how superstring theory provides a description of the
underlying degrees of freedom which give rise to this entropy.
However, the specific nature of these degrees of freedom will not
concern us here.  For our purposes, a coarse grained description of
these degrees of freedom, which we will call the stretched horizon
[\stu ], can be used.

A simple model of the stretched horizon can be constructed by
considering a set of quantum fields $\phi^A$ propagating within a
spherical shell of proper thickness $\varepsilon$ in the vicinity of
the horizon [\thooft ].  The field theory is explicitly cut off by
restricting to modes with momentum less than the Planck mass
$m_{\Planck} = G^{-1/2}$.  The fields within this shell are coupled
in some specific manner to the fields outside.  The field operator
can be written as
$$
\eqalign{\phi^A  &= \sum_{\ell = 0}^{\ell_{\MAX}}
\sum_{m=-\ell}^{\ell} \phi^A_{\ell, m} \>, \cr
\phi^A_{\ell, m} (t,r,\theta,\varphi) &= f^A_{\ell, m} (t,r)
\Omega_{\ell, m}(\theta,\varphi) \>, \cr}
\eqn\fsum
$$
where $\Omega_{\ell, m}$ is the appropriate spherical harmonic.
Since we have restricted to momentum modes less than the Planck mass,
the maximum allowed angular momentum is
$$
\ell_{MAX} \approx | {\vec L} | = | {\vec x} \times {\vec p}_{\MAX} |
= 2M \sqrt{G} \>,
\eqn\lmax
$$
and the total number of allowed angular momentum modes is
$$
N = \sum_{\ell = 0}^{\ell_{MAX}} (2\ell + 1) = (\ell_{\MAX} + 1)^2
\approx 4 M^2 G \propto {A \over {4 G}}\>.
\eqn\modes
$$
If we now treat the fields outside the stretched horizon as a heat
bath in thermal contact with the stretched horizon, the thermal
entropy of the stretched horizon is proportional to $N$, and thus is
proportional to ${A \over {4G}}$.  In other words, modes with angular
momentum up to $2M \sqrt{G}$ are important in accounting for the
Bekenstein-Hawking entropy \BHent .  If one truncates the system down
to the spherically symmetric modes $(\ell = 0)$, the above simple
analysis shows that the entropy should no longer be proportional to
the area of the black hole.  It should not be surprising that when
all of the degrees of freedom which could account for the entropy are
truncated, information is lost.

%
%
\chapter{Mirrors and the Origin of Hawking Radiation}

Next, we shall consider the effects of neglecting interactions.  To
this end, we will examine the following gedanken experiment.
Consider an evaporating Schwarzschild black hole of mass $M \gg
m_{\Planck}$.  Let us focus attention on a very unlikely event:
suppose the Hawking radiation assembles itself into a spherical,
perfectly reflecting mirror at a proper distance $\varepsilon$ above
the horizon.  This mirror reflects the outgoing radiation back into
the hole and any incoming radiation back out to infinity.\foot{Now,
we do not believe any more than you do that if $\varepsilon = {\cal
O} (\ell_{\Planck})$, that any physical mirror could withstand the
Planckian temperatures in this region.  We are really considering a
purely mathematical exercise involving a fixed, classical geometry
describing the black hole, and quantum fields propagating on that
geometry.  Mathematically, the mirror is a reflecting boundary
condition on the fields at the proper distance $\varepsilon$ from the
horizon.  In the context of this mathematical model, we are
interested in the consequences of nontrivial interactions between the
various angular momentum modes.}  The region outside the mirror can
then be studied as a system with a perfectly reflecting boundary
condition at the mirror.  The question we want to address is, for how
long will the system continue to radiate as seen from the outside?

We will first study the system using the approximation of free
fields.  Since the lifetime of the black hole is of order $G^2 M^3$,
for times small compared to $G^2 M^3$ we can approximate the exterior
geometry by the usual Schwarzschild geometry \linelement .  Before
the mirror appears, the state of the system is the Hartle-Hawking
vacuum.  An observer at fixed radial coordinate $r$ close to $2MG$
experiences an approximately thermal flux of Hawking particles with
proper temperature given by
$$
T = {1 \over {8 \pi MG \sqrt{1 - {{2MG} \over r}}}} \>,
\eqn\temp
$$
which can be approximated by
$$
T \approx {1 \over {2 \pi \rho}}
\eqn\temptwo
$$
where $\rho$ is the proper distance from the event horizon.  Thus an
observer near the horizon basically sees the Unruh thermal state.  An
observer far from the black hole, however, would not describe the
state as precisely thermal.  Because of the angular
momentum-dependent effective potential experienced by fields
propagating in the fixed Schwarzschild spacetime, almost all of the
Hawking radiation that reaches infinity is in low angular momentum
modes.

We now consider the theory of a free massless scalar field $\phi$.
The system is most easily analyzed if we change to the Regge-Wheeler
tortoise coordinate
$$
r_* = r + 2MG \log \left ( {r \over {2MG}} - 1 \right ) \>,
\eqn\rwtort
$$
for which the line element takes the conformal form
$$
ds^2 = \left ( 1 - {{2MG} \over {r}} \right ) \left [ -dt^2 + dr_*^2
\right ] + r^2 d\Omega^2 \>.
\eqn\tortlinel
$$
In these coordinates, the mirror surface is at $r_{* \> \mirror} =
2MG \left [ 2 \log \left ( {\varepsilon \over {4MG}} \right ) + 1 +
\left ( {\varepsilon \over {4MG}} \right )^2 \right ]$.  Writing the
field as
$$
\phi (t,r_* ,\theta,\varphi) = \sum_{\ell, m} \int_{-\infty}^{\infty}
{{dE} \over {(2\pi)}} e^{-iEt} {{U_{E \ell m} (r_*)} \over r}
Y_{\ell}^m (\theta,\varphi)
\eqn\fieldecomp
$$
the field equation for $U$ can be written as a time-independent
Schr\"odinger equation
$$
\left \{ -{{d^2} \over {dr_*^2}}  + V_{\eff}(r_*; \ell) \right \}
U_{E \ell m} = E^2 U_{E \ell m} \>,
\eqn\ueq
$$
where the effective radial potential is
$$
V_{\eff} = \left [ {{r - 2MG} \over r^3} \right ] \left [ \ell (\ell
+ 1) + {{2MG} \over r} \right ] \>.
\eqn\veff
$$
and $E$ is the energy of the mode as seen by an observer at infinity.
 $V_{\eff}$ has a global maximum at $r \approx {8 \over 3} MG$,
corresponding to the tortoise coordinate $r_{* \> \outer}$.  For the
case $\ell = 0$, the barrier height is ${9 \over {1024 (MG)^2}}$, and
it increases monotonically with $\ell$.  Since the great majority of
the populated modes have energy $E \approx T_H$, we see that only
very few of the higher angular momentum modes can tunnel through the
potential barrier and escape.

Now consider the effect of the appearance of the mirror.  The higher
angular momentum modes are effectively trapped between the mirror and
the potential barrier at $r_{* \> \outer}$.  Only the lowest angular
momentum modes can escape, and in the free field approximation they
are decoupled from the higher angular momentum modes.  For simplicity
in this discussion, we will drop all but the $s-$wave.  To calculate
the lifetime of the radiation, we first calculate how long it takes
an $s-$wave mode to propagate from the mirror surface to the outer
reaches of the black hole, which we define to be $r_{* \> \outer}$.
We then multiply the result by $2 e^{2A}$, where $e^{-A}$ is the
amplitude to tunnel through the barrier.  The amount of time it takes
for an $s-$wave to propagate from $r_{* \> \mirror}$ to $r_{* \>
\outer}$ is simply
$$
\delta t = r_{* \> \outer} - r_{* \> \mirror} = MG \left (4 \log
\left ( {{\sqrt{8} MG} \over \varepsilon} \right ) + 1 + {\cal O}
\left ( {{\varepsilon^2} \over {M^2 G^2}} \right ) \right ) \sim MG
\log \left ( {{MG} \over \varepsilon} \right ) \>.
\eqn\time
$$
It turns out that the tunneling suppression is independent of $M$ for
particles with energy of order $T_H$ and is ${\cal O}(1)$ for the
$s-$wave particles [\page ].

Now, in the absence of a mirror, the black hole radiates for a time
$t_{\evap} = {\cal O} (G^2 M^3)$.  The above calculation tells us
that the mirror will shut down the Hawking radiation after a time
much less than $t_{\evap}$ unless $\varepsilon$ is of order
$$
\varepsilon \sim MG \exp (-M^2 G) \>,
\eqn\toosmall
$$
which is an absurdly small distance and cannot be physically
meaningful.  If we restrict ourselves to distances larger than
$\ell_{\Planck}$, the majority of the Hawking radiation responsible
for the evaporation of the black hole is trapped behind the mirror.
The conclusion is that the Hawking radiation originates at distance
scales of the order given in equation \toosmall .  Another
consequence is that after a time of order $\delta t$, an external
observer will be able to see his own reflection in the mirror
surface.  The photons he emits will propagate freely down to the
mirror surface and reflect right back out.

The reasoning used above can be applied to determine where the
Hawking radiation originates even without the mirror.  For example,
starting from a black hole of mass $M$ and Hawking temperature $T_H =
{1 \over {8\pi MG}}$, suppose we let the black hole evaporate until
its new temperature is given by $T_H' = T_H (1 + \delta)$.  $\delta$
can be chosen arbitrarily small, say $10^{-6}$, so the approximation
of the system by the static Schwarzschild metric is a good one.  The
amount of Schwarzschild time $t$ needed for this process is
proportional to $\delta M^3 G^2$.  A calculation identical to the
above leads to the belief that the Hawking radiation which is
responsible for this evaporation originates at distances of the
absurdly small order given in equation \toosmall .

Now, let us return to the real world, in which interactions exist
between particles.  We will again consider the case of a mirror
appearing at proper distance $\varepsilon$ above the horizon.  The
region just outside the mirror is at proper temperature $T = {1 \over
{2\pi \varepsilon}}$.  The strength of the interactions in this
region is governed by the values of running coupling constants,
evaluated at momentum scales of order $T$.  These interactions couple
the different angular momentum modes, so that every now and then a
higher angular momentum particle will get scattered into the
$s-$wave, and may then escape.  Since the higher angular momentum
modes are essentially confined to the region between the mirror and
$r_{* \> \outer}$, the result is a slow replenishment of the
$s-$wave, which allows the system to continue radiating much longer
than one would expect from the naive free field calculation.  Let us
estimate how often an $s-$wave quantum is produced.

Consider a thin spherical shell at radial coordinate $r$.  Let the
proper distance from the shell to the horizon be $\rho (r)$, and let
the proper thickness of the shell be $\Delta \rho (r)$.  Most of the
particles in the shell will have momentum of the order of the proper
temperature of the shell,
$$
T(r) = {1 \over {8 \pi MG \sqrt{1 - {{2MG} \over r}}}} \approx {1
\over {2\pi \rho (r)}} \>.
\eqn\proptemp
$$
Since most of the particles have momentum of order $T$, it makes no
sense to choose $\Delta \rho (r)$ much smaller than ${1 \over T}$.
Further, since the temperature is varying relatively rapidly, it also
makes no sense to choose $\Delta \rho (r)$ much bigger than $\rho$.
We will therefore choose $\Delta \rho (r) = \rho (r)$.  We can then
treat the shell as an interacting neutral plasma at proper
temperature $T(r)$.  By dimensional analysis, the number of
collisions per unit proper time per unit proper volume due to a
particular interaction will have the form
$$
{{dn} \over {dVd\tau}} = \alpha^2(T) T^4 \>,
\eqn\colleqn
$$
where $\alpha$ is an average dimensionless running coupling constant
evaluated at the momentum scale $T$.  The proper volume of the shell
is $dV = 4\pi r^2 \Delta \rho$, so the number of collisons per unit
proper time in this shell is
$$
{{dn} \over {d\tau}} = 4\pi \alpha^2 T^4 r^2 \Delta \rho \>.
\eqn\neqn
$$

Since most of the particles have momentum of order $T$, only those
modes with angular momentum less than
$$
\ell_{MAX} \approx | {\vec L} | = | {\vec x} \times {\vec p} | = rT
\>,
\eqn\lmax
$$
can be relevant in either the initial or the final state of a
collision.  The total number of such modes is
$$
N = \sum_{\ell = 0}^{\ell_{MAX}} (2\ell + 1) = (\ell_{MAX} + 1)^2
\approx  (rT)^2 \>.
\eqn\modes
$$
When a typical pair of particles in the plasma collide, the
probability that one of them is scattered into any given angular
momentum mode is of order ${1 \over N}$.  Thus, the number of
$s-$wave particles produced per unit proper time in the shell is
$$
{{dn_s} \over {d\tau}} \approx {1 \over N}{{dn} \over {d\tau}} = 4\pi
\alpha^2 T^2 \Delta \rho \>.
\eqn\sproduce
$$
In order to compute the number of $s-$wave particles produced per
unit Schwarzschild time, we simply need to multiply equation
\sproduce\ by the redshift factor ${{d\tau} \over {dt}}$.  Inserting
the expression \proptemp\ for the proper temperature, we obtain
$$
{{dn_s} \over {dt}} \approx {{\alpha^2 (r) \Delta \rho (r)} \over {16
\pi G^2 M^2 \sqrt{1 - {{2MG} \over 4}}}} \>.
\eqn\dndt
$$
This expression must be summed over shells, starting with the shell
which begins at the mirror surface, at proper distance $\varepsilon$
from the horizon.  At temperatures below the mass of the electron,
all cross sections go rapidly to zero, so the last shell to be
include in the sum should have temperature of order $m_e$.  The sum
can be approximated by an integral
$$
{{dn_s} \over {dt}} \approx {1 \over {16\pi G^2 M^2}}
\int_{\varepsilon}^{{1 \over {2\pi m_e}}} d\rho {{\alpha^2} \over
{\sqrt{1 - {{2MG} \over r(\rho)}}}} \>.
\eqn\intone
$$
Since $\alpha$ can vary no more rapidly than a logarithm, we replace
$\alpha(\rho)$ by an average ${\overline \alpha}$, and the integral
is easily evaluated.  Dropping all but the leading behavior, we find
$$
{{dn_s} \over {dt}} \sim {{{\overline \alpha}^2 \log \left ( {1 \over
{m_e \varepsilon}} \right )} \over {4 MG}} \>.
\eqn\dnsdt
$$
This is to be compared to the ordinary flux of particles as seen by
an observer at infinity,
$$
{{dn_{\hbox{\sevenrm Hawking}} \over {dt}} \sim {1 \over {4MG}}} \>.
\eqn\dnsdthawk
$$
Thus we see that the maximum replenishment rate is of the order of
${\overline \alpha}^2 \log \left ( {1 \over {m_e \varepsilon}} \right
)$ times the Hawking rate.  If the mirror appeared at a GUT distance
above the horizon, the rate of replenishment would be insufficient to
sustain the Hawking emission rate.  However, the black hole would
continue to radiate at a diminished rate until all of the particles
in the thermal atmosphere above the mirror were depleted.  Since the
number of particles is approximately given by the entropy \BHent , we
expect the black hole to radiate until a time of order
$$
t \sim {{M^3 G^2} \over {{\overline \alpha}^2 \log \left ( {1 \over
{m_e \varepsilon}} \right )}} \>.
\eqn\tnew
$$

On the other hand, the mirror need only be at a distance of order the
Planck length to make gravitational interactions strong enough to
sustain the Hawking radiation fully.  In this case, no noticeable
effect of the mirror could be discerned for a time of order $G^2
M^3$.  This means that the true origin of the Hawking radiation is at
distance of the order of the Planck length from the horizon.  There
is no need to invoke distances of order $MG \exp (-M^2 G^2)$.

Regardless of where the mirror occurs, an external observer would not
be able to see his reflection in the mirror until after a time of
order $G^2 M^3$, if at all.  As long as the thermal atmosphere of the
black hole remained, the photons he emitted would be scattered and
thermalized near the horizon.  It would be expected that the
information carried by these photons would not be radiated back out
until a time of order $G^2 M^3$.

This example raises the interesting question of whether it is at all
possible to detect the presence of a mathematical mirror located at a
Planck distance from the horizon of an evaporating black hole.  `t
Hooft has speculated that such a mirror would, in fact, be
undetectable [\thooft ].

%
%
\chapter{Particles and Gauges}

An argument which is often raised about calculations of the type
presented above goes as follows.  Consider an infalling observer and
a stationary observer who stays permanently outside the black hole,
both near the horizon.  As mentioned previously, a stationary
observer near the horizon sees a thermal bath of particles at proper
temperature given in equation \temp .  For a sufficiently massive
black hole, however, the infalling observer does not see the hot bath
of particles near the horizon, since he can perform no local
experiment to detect the presence of the horizon.  Because the
infalling observer does not see the thermal bath, it is claimed that
it is not a physical phenomenon--only the $s-$wave particles observed
at distances greater than $r=3MG$, the existence of which both a
stationary and an infalling observer will agree upon, are physical.
Therefore, it is claimed, the replenishment of the $s-$wave modes
calculated above cannot be physical, and one is back to discussing
absurdly short distances.

The fallacy of this argument comes from not being true to one's
choice of gauge, {\it i.e., \/} of one's coordinate chart.  The
infalling and stationary observers describe physics in different
gauges.  It makes no sense to dismiss the description of a system
made in one gauge as unphysical, while claiming that the description
made in another is physical.  Within a given gauge, the only
criterion for the physicality of phenomena is that the theory be
internally consistent.  This means, in particular, that the
stationary observer, who uses coordinates covering only the region
outside the black hole, has no choice but to include all mathematical
degrees of freedom that are required for a consistent description in
his coordinate system.  One is not allowed to throw away degrees of
freedom in one gauge because they do not appear in another gauge.

As an example, consider quantum electrodynamics.  If one chooses to
perform calculations using the Coulomb gauge, one finds the existence
of long range instantaneous interactions, but the Hilbert space of
states is manifestly positive definite.  If one instead chooses to
work in Lorentz gauge, one no longer finds long range instantaneous
interactions, but the Hilbert space one works in now contains
unphysical states containing longitudinally and timelike polarized
photons.

Throwing away the long range instantaneous interaction in Coulomb
gauge, because it is not present in Lorentz gauge, is clearly a
serious mistake.  Quantum electrodynamics in the Coulomb gauge
requires the existence of the long range instantaneous interaction
for mathematical consistency.  Likewise, throwing the longitudinal
degrees of freedom in Lorentz gauge because they are not present in
the Coulomb gauge destroys the internal consistency of the theory.
It is an equally foolish argument to drop the effects of the high
energy, high angular momentum thermal atmosphere in the stationary
coordinate system because it is absent in the infalling coordinate
system.

The description of the black hole in each coordinate system
separately appears to be internally consistent.  The infalling
observer falls freely past the horizon, never seeing any high energy
thermal bath, but can never communicate this information to the
observer outside.  The outside observer sees a hot thermal bath and
sees the infalling observer disappear into it.  The confusion arises
when one tries to relate the two descriptions.  As long as the
infalling observer remains outside the black hole, however, there
exists a gauge transformation which will map his local description of
physics into that given by the stationary observer.  The gauge
transformation is simply the coordinate transformation between the
two frames of reference.

To be explicit, let us consider a region very near the horizon, which
is approximated by Rindler space.  Rindler space is simply the
section of Minkowski space as seen by a uniformly accelerated
observer, and freely falling particles move on straight lines in
Minkowski space.  The coordinate transformation between Rindler
coordinates $(t, \rho, y, z)$ and Minkowski coordinates $\{ x^{\mu}
\}$ is given by
$$
\eqalign{x^0 &= \rho \sinh \left ( {t \over {4MG}} \right ) \>, \cr
x^1 &= \rho \cosh \left ( {t \over {4MG}} \right ) \>, \cr}
\eqn\rind
$$
The horizon is at $t = \infty$.  The effect of a time translation in
the Rindler time is equivalent to a boost in Minkowski space.  Now
consider a particle falling toward the horizon.  The relation between
the rest frame of a freely falling particle and that of a stationary
Rindler observer is given by a time-dependent boost angle which
increases to infinity as the particle approaches the horizon.  This
means that to relate the descriptions of physics in the two frames
requires a knowledge of how physical states behave under extremely
large Lorentz boosts.  The boost angle becomes as large as $M^2 G$
during the lifetime of the black hole.  It therefore follows that the
Rindler momentum of the particle becomes as large as $e^{M^2 G}$.
The black hole is the ultimate particle accelerator.  In trying to
formulate the relation between the physics in these two coordinate
frames, we are driven to a range of relative momenta with which we
have little experience.

This leads us to the following hypothesis.  For a theory to be
consistent with the existence of black holes, it must be such that
the effect of a super-Planckian Lorentz boost on a system is
equivalent to the accumulated effect of putting the system into
contact with a thermal bath at Planckian temperature for a long
period of time.  We might therefore expect that under a sufficiently
large boost, a particle appears to melt and diffuse over a large
region of space.

When a freely falling observer passes the horizon, there is no longer
any coordinate transformation which will map his local description of
physics to that of the stationary observer.  This is because the
stationary observer can never measure what goes on behind the
horizon.  It is in this sense that measurements made by the two
observers are {\it complementary.} [\stu ]

%
%
\chapter{Superstrings and Black Holes}

The hypotheses given above give us some idea of how matter should
behave under large Lorentz boosts in a quantum theory of gravity
containing black holes.  Now we should look for ways to implement
this idea.  It will be seen that superstring theory exhibits the
properties listed above.  In the previous sections we assumed that
the standard laws of physics hold down to the Planck scale.  In
string theory however, the new physics begins at the {\it string
scale} which differs from the Planck scale by factors of the
dimensionless string coupling constant $\kappa$.  If $\ell_{\Planck}
= \sqrt{G}$ is the Planck length and $\ell$ is the string length then
$$
\ell_{\Planck} = \kappa \ell \>,
\eqn\slength
$$
so that if $\kappa$ is very small the new physics begins at length
scales appreciably larger than $\ell_{\Planck}$.  In what follows we
will use units in which $\ell = 1$.  From what we have argued, the
environment sufficiently near the horizon as described in
Schwarzschild coordinates should resemble the phase of string theory
above the Hagedorn temperature.  We expect this to consist of a
condensate or very dense hot soup of strings strongly interacting
with each other [\witat , \suss , \su ].  Furthermore, it has been
argued [\suss, \su, \dab ] that the Bekenstein-Hawking entropy per
unit area of a horizon can be understood as the entropy of this soup
of strings.

Let us proceed to consider the description of strings by a stationary
Schwarzschild observer.  The propagation of closed superstrings in a
Schwarzschild background has not been completely analyzed.  If we use
the Rindler space approximation of the near horizon geometry,
however, then we can use results from flat space string theory.

The points of Minkowski space occupied by a string are given by
functions $X^{\mu}(\tau, \sigma)$, where $\tau$ and $\sigma$ are
coordinates on the string world sheet.  In addition, the string has
internal degrees of freedom implied by supersymmetry and
compactification.  In  the light cone frame, $X^+ = (X^0 +
X^1)/\sqrt{2} = \tau$, and the dynamical degrees of freedom are the
two transverse coordinates $\{ X^i \}_{i=1}^2$, which are decoupled
from the internal degrees of freedom.  The normal mode decomposition
of $X^i$ for a noninteracting superstring is the same as for a free
bosonic string
$$
X^i (\tau, \sigma) = x^i + p^i \tau + {i \over 2} \sum_{n \ne 0} {1
\over n} \left [ \alpha^i_n e^{-2in(\tau - \sigma )} + \alphat^i_n
e^{-2in(\tau + \sigma )} \right ] \>,
\eqn\boson
$$
where $x$ and $p$ are the center of mass position and momentum, and
the $\alpha$ are the mode coefficients.

The transverse size of the string at light cone time $\tau = 0$ can
be estimated by computing the expectation value of
$$
\Rvec^2 = {1 \over \pi} \int d\sigma \left ( \Xvec (\sigma) - \xvec
\right )^2
\eqn\tsize
$$
in whatever state is under consideration.  Now if an observer uses an
apparatus with resolution time $\varepsilon$ to measure the size of
the string, he should only include in his description of the string
modes with frequency less than ${1 \over \varepsilon}$ in his frame
of reference.  The frequency of mode $n$ is given by
$$
\nu_n = {n \over {P}} \>,
\eqn\freq
$$
where $P$ is the longitudinal momentum of the string as measured by
the observer.  This means that we need only include modes with $|n|
\le N = P/\varepsilon$.  For the ground state one easily finds
$$
\bra{0} \Rvec^2 \ket{0} = \sum_{n = 1}^N {1\over n} \approx \log
\left ( {P \over \varepsilon} \right ) \>.
\eqn\gstate
$$
Here we see the first example of anomalous behavior of strings under
Lorentz boosts.  Instead of the transverse size being independent of
momentum, one finds that it depends on the momentum logarithmically.
A more complete analysis of the growth of the transverse size of free
strings was made in [\kks ], and indicates that in addition the
length of string in the transverse plane is proportional to $P$.  As
$P$ increases, the string loops back over itself many times, in such
a way that the total area occupied by the string only grows
logarithmically.  As the momentum goes to infinity, the string
becomes dense over all of space.  In Figures 1 (a) to 1 (f), we show
a sequence of snapshots of a string falling toward a Rindler horizon,
taken by a stationary Rindler observer at equal intervals of Rindler
time.  The figures were generated by imposing a smooth mode cutoff on
a string wave function, a procedure very similar to that used in
[\kks ].  In figure 1 (a), only the very lowest modes appear, but as
Rindler time progresses, more and more modes enter the description.

Even the modest logarithmic growth of a free string has surprising
implications to a Schwarzschild observer.  Recalling that the
longitudinal momentum of a string falling toward the horizon as
measured by the Schwarzschild observer grows exponentially with time,
we see that the area of the region occupied by the string grows like
$\vev{\Rvec^2} = t/4MG$.  This behavior has been successfully
interpreted as an effective thermalization of the string, causing it
to melt and diffuse over the horizon [\mpt ].  If no other effects
take place the string would grow to a size comparable to the
Schwarzschild radius in a time of order $\kappa^2 G^2 M^3$.  If
$\kappa$ is small, this time is short in comparison with the
evaporation time of the black hole.

The process is very similar to the stochastic evolution of a scalar
field in an inflating universe.  In both cases more and more modes
enter the description with time.  These modes enter with random
phase and amplitude.  In each case the growth and spreading can be
described by stochastic interactions with a heat bath.  In the string
case the heat bath is provided by the Unruh effect.  Thus we see that
the second of our hypotheses is realized even at the level of free
string theory.

Another suprise to the Schwarzschild observer is that the string
fails to Lorentz contract as the momentum gets large.  This is
important for the finiteness of the entropy of the horizon.  The
ordinary view is that particles Lorentz contract along the direction
of their motion by a factor proportional to ${1 \over P}$.  For this
reason, an arbitrarily large number of particles can be stacked near
the horizon [\mempar ].  This is in obvious contradiction to the
finiteness of the Bekenstein-Hawking entropy (and the possibility of
the existence of a mirror).  The longitudinal behavior of strings is
quite different.  To compute the mean longitudinal spread $\Delta
X^-$ at $\tau = 0$ we use the constraint equation [\ggrt ]
$$
{{\pa X^-} \over {\pa \sigma}} = {{\pa \Xvec} \over {\pa \sigma}}
\cdot {{\pa \Xvec} \over {\pa \tau}} + I
\;,
\eqn\constrnt
$$
where $I$ represents the contribution from compactified modes,
fermionic modes, {\it etc.}  We can rewrite equation \constrnt\ in
terms of the transverse Virasoro generators, obtaining
$$
{{\pa X^-} \over {\pa \sigma}} = \sum_{n \ne 0} \left ( L_n e^{in
\sigma} - {\tilde L}_n e^{-in \sigma} \right ) \>,
\eqn\virasoro
$$
which can be integrated to give
$$
X^-(\sigma) = x^- + \sum_{n \ne 0} {1 \over {in}} \left [ L_n e^{in
\sigma} + {\tilde L}_n e^{-in \sigma} \right ] \>.
\eqn\xminus
$$
Using the standard Virasoro algebra one finds
$$
\bra{0} \left ( \Delta X^- \right )^2 \ket{0} \approx 4 \sum_{n=1}^N
n \approx 2 \left ( {P \over \varepsilon} \right )^2 \>,
\eqn\nocont
$$

Equation \nocont\ indicates that no Lorentz contraction of the string
distribution takes place.
The spreading process begins to occur when the string reaches a
distance of order the string scale from the horizon.  The result is
that in the Schwarzschild coordinates the bulk of the string never
approaches closer than a distance of order $\ell_{\Planck}$ to the
horizon.  This peculiar property of strings supports our first
hypothesis that an external observer cannot detect the presence (or
absence!) of a mirror at a distance of order $\ell_{\Planck}$ from
the horizon.

Eventally, the density of string will become so large that
interactions can no longer be ignored, and more complicated phenomena
occur, which probably cause the transverse growth to become even more
rapid.  As the string replicates, its transverse density increases.
At the center of the distribution the average number of strings $N$
passing through a region of area $A$ is of order $\exp ( \Rvec^2 )
\sim \exp \left ({t \over {4M}} \right )$.  However, this enormous
density of string certainly leads to new effects once it becomes of
order ${1 \over \kappa^2}$.  At this time the probability for string
interactions becomes unity and perturbation theory breaks down.  One
attractive possibility is that the growth of string density is cut
off at this point.  The result would be that the density grows until
there is about one string per unit Planck area.  This is also
suggested by the fact that the entropy of a black hole is
proportional to its area [\susstwo ].

Perhaps the most remarkable aspect of the above description is that
none of it is seen by an observer who falls through the horizon with
the string.  Such an observer sees the string with a fixed time
resolution and therefore sees a constant transverse and longitudinal
size as the horizon is crossed.

To conclude, we would like to point out some questions which need to
be addressed.  To begin, we must understand the behavior of the hot
string soup at the horizon, both with and without a mirror present.
Our speculation is that the presence of a mirror at a distance of
about the Planck scale from the horizon will not affect the answer.
Not unrelated to this, the behavior of string interactions when the
density of string gets to be of order $\kappa^{-2}$ needs to be
understood.

%
%
\ack

We would like to thank R.~Sorkin for a discussion which led to the
considerations in the first part of this paper, and the organizers
and participants of the 1994 PASCOS conference for providing a
stimulating meeting.  This work is supported in part by National
Science Foundation grant PHY89-17438.  J.~U. is supported in part by
a National Science Foundation Graduate Fellowship.

\refout

\vfill\endpage

\centerline{\bf FIGURE CAPTIONS}

Figure 1 (a) - 1 (f): Snapshots of a string falling toward a Rindler
horizon, taken by a stationary Rindler observer at equal increments
of Rindler time.  In Figure 1 (a), only the lowest modes contribute
to the effective string wave functional, but as time progresses, more
modes enter the description.  Figure 1 (f) shows many modes have now
entered the effective wave functional.  We see the roughly linear
growth of the area, and that the density of string near the center of
the distribution is getting very large.

\end